\documentclass[table,sigconf]{acmart}

\usepackage{amsmath,amsfonts}
\usepackage{array}
\usepackage{algorithmic}
\usepackage{graphicx}
\usepackage{textcomp}
\usepackage{xcolor}
\usepackage{float}
\usepackage{booktabs}
\usepackage{subcaption}
\usepackage{multirow}
\usepackage{enumitem}
\usepackage{tabularx}
\usepackage{adjustbox}
\usepackage[most]{tcolorbox}
\usepackage{balance}
\usepackage{hyperref}
\usepackage{listings}
\usepackage{tipa}
\usepackage{svg}
\usepackage{color, colortbl}
\usepackage{soul}
\usepackage{fancyhdr}
\usepackage{subfloat}
\usepackage[skip=1pt,labelfont=bf]{caption}
\usepackage{marginnote}
\usepackage[colorinlistoftodos]{todonotes}
\usepackage{cancel}
\copyrightyear{2024}
\acmYear{2024}
\setcopyright{rightsretained}
\acmConference[ICSE '24]{2024 IEEE/ACM 46th International Conference on Software Engineering}{April 14--20, 2024}{Lisbon, Portugal}
\acmBooktitle{2024 IEEE/ACM 46th International Conference on Software Engineering (ICSE '24), April 14--20, 2024, Lisbon, Portugal}\acmDOI{10.1145/3597503.3639074}
\acmISBN{979-8-4007-0217-4/24/04}

\begin{document}

\newcommand\ans[1]{
\noindent
\fcolorbox{green!40!black}{green!5}{\noindent
 \parbox{0.98\columnwidth}{\noindent  #1}}\\
}

\title{Unveiling Memorization in Code Models}

\definecolor{codegreen}{rgb}{0,0.6,0}
\definecolor{codegray}{rgb}{0.5,0.5,0.5}
\definecolor{codepurple}{rgb}{0.58,0,0.82}
\definecolor{backcolour}{rgb}{0.95,0.95,0.92}

\lstset{
   xleftmargin=2em,
   xrightmargin=0em,
}

\lstdefinestyle{mystyle}{
  language=Python,
  backgroundcolor=\color{backcolour},
  basicstyle=\small\ttfamily,
  numbers=left,
  numberstyle=\tiny\color{codegray},
  stepnumber=1,
  numbersep=10pt,
  tabsize=4,
  showspaces=false,
  showstringspaces=false,
  showtabs=false,
  frame=single,
  rulecolor=\color{black},
  aboveskip=0pt,
  belowskip=0pt,
  captionpos=b,
  stringstyle=\color{codepurple},
  keywordstyle=\color{blue},
  commentstyle=\color{codegreen},
  escapeinside={\%*}{*)},
  morekeywords={*,...}.
}

\lstset{style=mystyle}

\newboolean{showcomments}
\setboolean{showcomments}{true}
\ifthenelse{\boolean{showcomments}}
{ \newcommand{\mynote}[2]{\textcolor{orange}{
			\fbox{\bfseries\sffamily\scriptsize#1}
			{\small$\blacktriangleright$\textsf{\emph{#2}}$\blacktriangleleft$}}}}

\ifthenelse{\boolean{showcomments}}
{ \newcommand{\dnote}[2]{\textcolor{red}{
			\fbox{\bfseries\sffamily\scriptsize#1}
			{\small$\blacktriangleright$\textsf{\emph{#2}}$\blacktriangleleft$}}}}

\ifthenelse{\boolean{showcomments}}
{ \newcommand{\hnote}[2]{\textcolor{blue}{
			\fbox{\bfseries\sffamily\scriptsize#1}
			{\small$\blacktriangleright$\textsf{\emph{#2}}$\blacktriangleleft$}}}}

\newcommand{\mh}[1]{\hnote{Hilmi}{#1}}
\newcommand{\dl}[1]{\dnote{David}{#1}}
\newcommand{\dongsun}[1]{\dnote{Dongsun}{#1}}
\newcommand{\yz}[1]{\dnote{Zhou}{#1}}
\newcommand{\hjd}[1]{\dnote{Junda}{#1}}

\author{Zhou Yang\textsuperscript{$\blacklozenge$}, Zhipeng Zhao\textsuperscript{$\varheartsuit$}, Chenyu Wang\textsuperscript{$\blacklozenge$}, Jieke Shi\textsuperscript{$\blacklozenge$},  \\ Dongsun Kim\textsuperscript{$\spadesuit$}, Donggyun Han\textsuperscript{$\clubsuit$}, and David Lo\textsuperscript{$\blacklozenge$}}

\affiliation{%
  \institution{\textsuperscript{$\blacklozenge$}School of Computing and Information Systems, Singapore Management University, Singapore}\country{}
}
\affiliation{%
  \institution{\textsuperscript{$\varheartsuit$}Department of Computer Science, University of Copenhagen, Copenhagen, Denmark}\country{}
}
\affiliation{%
  \institution{\textsuperscript{$\spadesuit$}School of Computer Science and Engineering, Kyungpook National University, Daegu, South Korea}\country{}
}
\affiliation{%
  \institution{\textsuperscript{$\clubsuit$}Department of Computer Science, Royal Holloway, University of London, London, UK}\country{}
}
\affiliation{%
  \institution{\{zyang, chenyuwang, jiekeshi, davidlo\}@smu.edu.sg, zpzhao.zzp@gmail.com, darkrsw@knu.ac.kr, DongGyun.Han@rhul.ac.uk}\country{}
}








\renewcommand{\shortauthors}{Zhou Yang, Zhipeng Zhao, Chenyu Wang, Jieke Shi, Dongsun Kim, DongGyun Han, David Lo}

\begin{abstract}

The availability of large-scale datasets, advanced architectures, and
powerful computational resources have led to effective code models that automate diverse software engineering activities.
The datasets usually consist of billions of lines of code from both open-source
and private repositories. 
A code model memorizes and produces source code
verbatim, which potentially contains vulnerabilities, sensitive information, or code with strict licenses, leading to potential security and privacy
issues.

This paper investigates an important problem: \textit{to what extent do code models \underline{memorize} their training data?}
We conduct an empirical study to explore memorization in large pre-trained code models.
Our study highlights that simply extracting 20,000 outputs (each having 512 tokens) from a code model can
produce over 40,125 code snippets that are memorized from the training data.
To provide a better understanding, we build a taxonomy of memorized contents 
with 3 categories and 14 subcategories.
The results show that the prompts sent to the code models affect the
distribution of memorized contents.
We identify several key factors of memorization. 
Specifically, given the same architecture, larger models suffer more from memorization problem.
A code model produces more memorization when it is allowed to generate longer outputs.
We also find a strong positive correlation between the number of an output's
occurrences in the training data and that in the generated outputs, which
indicates that a potential way to reduce memorization is to remove duplicates in the training data.
We then identify effective metrics that infer whether an output contains memorization accurately.
We also make suggestions to deal with memorization.

\end{abstract}



\maketitle

\section{Introduction}

As more large open-source datasets become publicly
available~\cite{husain2019codesearchnet,pile}, code
models~\cite{codex,wang2021codet5,codegen,fried2023incoder,santacoder},
trained on billion lines of code, are now an important part of
software engineering.
The models automate a series of critical tasks such
as defect prediction~\cite{steenhoek2023empirical},
code review~\cite{CodeReviewer}, code generation~\cite{plbart} and software questions analysis~\cite{TechSumBot,PTM4TAG,he2023representation}.
These models have gone beyond academic exploration and have been widely deployed and used by a large number of users.
For example, GitHub \textsc{CoPilot}~\cite{github-copilot}, powered by the OpenAI
Codex model~\cite{codex}, obtains over 400,000 subscribers in the first month
of its release~\cite{torres_2022} and has already been used by over 1.2 million developers.
In addition, the recently released models~\cite{chatgpt,starcoder,fried2023incoder} demonstrate outstanding performance in software engineering tasks~\cite{chatgpt-repair,chatgpt-test-library}
and has powered many tools, e.g., IDE plugins.

The impressive performance of these models can be attributed to the combination
of advanced model architectures (e.g., state-of-the-art Transformer models with
Despite the remarkable advancements,
they are confronted with a set of privacy and legal challenges.
For example, \textsc{CoPilot} was found to produce real people's names and physical
addresses in its outputs~\cite{ding_2021}.
It is also reported that Samsung employees accidentally leaked company secrets
via \textsc{ChatGPT}~\cite{business_today_2023},
which raises concerns as \textsc{ChatGPT} retains records of these
conversations and could potentially use this data for training its system~\cite{openai_help_center}.

The above concerns emphasize the importance of \textit{exploring the capacity of large code models to memorize their training data}.
A code model memorizes a string from its training data if there exists a prompt to make the model generate this string.
This exploration is critical, especially given the fact that the
training data for these models may come from a variety of sources.
For example, large public datasets, such as the repositories hosted on GitHub,
are made publicly available for training code models.
These datasets may contain licensed code.
If code models memorize and generate licensed code, and model users are not
aware of this and utilize the code, it potentially results in a breach of
the license agreements.

The training data includes `software
secrets'~\cite{basak2023secretbench} such as passwords and API keys, which can
be outputted to users and leveraged by attackers directly.
For instance, as shown in our evaluation, a code model can memorize the username and password to access a database in a public IP address.
Furthermore, the training datasets may contain vulnerable or even malicious code.
Such code could potentially be memorized and displayed to users, causing a
serious risk to the security and integrity of software systems that adopt outputs of the code models.
Therefore, it is crucial to understand the memorization in code models.


To investigate the memorization of code models,
we first explore two open-source models: \texttt{CodeParrot} and \texttt{CodeParrot-small}~\cite{codeparrot}.
\texttt{CodeParrot} is a GPT-2 model with 1.5 billion parameters trained to generate Python code.
\texttt{CodeParrot-small} is a lightweight version of \texttt{CodeParrot} with 110 million parameters.
We choose to analyze the two models because (1) their training data is available and is of a size feasible to conduct memorization analysis;
(2) the models share the same architecture and dataset but come in two sizes, allowing us to investigate the impact of model size on memorization;

Our investigation begins by extracting a large number of outputs using different methods.
We extract 20,000 outputs by feeding a special token called the `start token' as a prompt into \texttt{CodeParrot};
this method provides no information to guide the generation.
After identifying over 40,125 unique memorized code snippets using clone detection, we conduct an open card sorting study on a statistically representative number (381) of memorized contents to build a taxonomy with 3 categories and 14 subcategories.
We find that 277 out of 381 memorized code snippets contain documentation (e.g., license information and docstring), and 239 out of 381 memorized code snippets contain code logic.

Additionally, we investigate factors that affect the memorization.
The results show that \texttt{CodeParrot} memorizes more contents than \texttt{CodeParrot-small}, indicating that larger models have stronger memorization ability.
Allowing a model to generate longer outputs can reveal more memorization.
When the maximal length of outputs increases, the number of memorized contents increases as well.
We also find a strong positive correlation between the number of a memorized content's
occurrences in the training data and that in the outputs,
which indicates that a potential way to reduce memorization is to remove duplicates in the training data.

Inspired by a study on memorization in language models~\cite{carlini21extracting}, we investigate four metrics to infer whether an output contains memorization: (1) perplexity~\cite{ppl}, (2) ratio of perplexity of two models, (3) ratio of perplexity to zlib~\cite{zlibnet}, and (4) average perplexity of sliding windows.
We compute these metrics to rank the outputs and analyze the top-100 ranking outputs.
We find that the first 3 metrics are very effective in identifying memorized contents: over 97\% of the top-100 outputs contain memorization.
However, metrics (1) and (3) tend to rank outputs with license information higher, while metric (2) tends to rank outputs with code logic higher.
Additionally, to demonstrate that memorization also exists in other code models, we analyze outputs from two popular code models that have already been deployed in practice: \texttt{Incoder}~\cite{fried2023incoder} and \texttt{StarCoder}~\cite{starcoder}.
We manually analyze the top 100 outputs from the two models ranked using metric (2).
We search code snippets in the top 100 outputs on GitHub to see whether GitHub repositories contain the same code.
If yes, it is an evidence that the model potentially memorize the code from its training data collected from GitHub.
We find that 81\% and 75\% of the top-100 outputs from \texttt{Incoder} and \texttt{StarCoder} contains code snippets that can be found on GitHub, respectively.
Besides, 90.12\% and 65.33\% of memorization are related to code logic.

To summarize, this paper makes the following contributions:
\begin{itemize}[leftmargin=0.5cm]
    \item We systematically investigate the phenomena of memorization in large code models, highlighting the potential risks of memorization in code models.
    \item We empirically show the feasibility to extract a large number of memorized contents from a code model using simple methods. We categorize the memorized contents into 14 categories and analyze their distribution.
    \item We analyze the factors affecting memorization in code models. Also, we evaluate metrics to infer memorization.
    \item We show that deployed code models memorize training data as well. We also provide some suggestions on how to deal with potential risks brought by such memorization.
\end{itemize}

\vspace*{0.2cm}
\noindent \textbf{Paper structure.}
Section~\ref{sec:background} provides the background and motivation.
In Section~\ref{sec:methodology}, we detail our methodology for extracting, identifying and inferring memorization.
Section~\ref{sec:settings} describes the experiment settings.
We analyze the memorization by answering three research questions in Section~\ref{sec:result}.
In Section~\ref{sec:discussion}, we provide some suggestions and discussion
Next, Section~\ref{sec:rel_work} highlights relevant studies.
We conclude our paper and present future work in Section~\ref{sec:conclusion}.

\section{Background and Motivation}
\label{sec:background}

This section describes the code generation process and some motivating examples to study such memorization.

\subsection{Code Generation Process}
\label{subsec:parameters}

Modern code models typically utilize the Transformer architecture~\cite{vaswani2017attention}.
Many well-known models (such as InCoder~\cite{fried2023incoder} and
codeparrot~\cite{codeparrot}) are built on the {GPT-2} model~\cite{gpt-2}, which is
referred to as the \textit{generative pre-trained transformer} model.
The primary objective of such models is to generate code based on a given context (i.e., the \textit{prompt} sent into the model's input layer).
This is accomplished by training the model on a large corpus of datasets and letting the model learn a probability distribution that predicts the likelihood of the next token, given the context.


Let us assume that the context consists of a sequence of tokens, denoted by $p = \langle x_1, x_2, \cdots, x_n \rangle$.
The model applies a softmax layer to compute the probability distribution of the next token, denoted by $P(x_{n+1} | p)$.
Formally, this distribution is defined:
\begin{equation}
	P(x_{n+1}|p) = \frac{e^{z_{n+1}}}{\sum_{i=1}^k e^{z_{i}}}
  \label{eq:softmax}
\end{equation}
In the equation, $z_{i}$ is the logit of the $i$-th token computed by the model.
Then, the model uses a \textit{decoding strategy} to decide the next token.
A commonly adopted strategy is the top-$k$ sampling~\cite{bengio2015scheduled}.
The model selects the $k$ most probable tokens from the probability distribution, and then chooses the next token from this new set of tokens.
The parameter $k$ plays an important role in balancing the trade-off between diversity and coherence in the generated output.
Intuitively, a small $k$ encourages the model to generate more fixed and coherent outputs, while a large $k$ introduces randomness and diversity.
This paper assumes users can control the code generation by modifying the $k$ parameter, which is feasible even with only access to the API (e.g., OpenAI API allows $k$ to be changed).

\subsection{Memorization Definition}
\label{subsec:definition}

Memorization in the context of generation tasks refers to a model's ability to store and recall specific information or patterns it has encountered during its training.
We define a string $s$ as a piece of memorization as follows.
Let $f: P \rightarrow S$ be a code model, where $P$ is the space of prompts and $S$ is the space of strings. Let $D \subseteq S$ be the training dataset. A string $s \in S$ is considered memorized if:
\begin{equation}
    \exists p \in P; s \in f(p) \land s \in D
\end{equation}
In other word, a string $s$ is memorized if there exists a prompt $p$ such that the model generates $f(p)$ when given $p$ as input, and $s$ appears both in the training dataset $D$ and the generated output $f(p)$.
In practice, the length of $s$ should be substantially long. Otherwise,
there might be many less useful memorized code such as variable names or boilerplate statements.
We explain the threshold to decide memorization from code models in Section~\ref{subsec:detection}.

\subsection{Memorization in Code Models}
\label{subsec:memorization}
The issue of memorization is especially concerning and should warrant careful attention from the research community.
We explain three scenarios where memorization may negatively impact the code models.
Each scenario comes with an example of memorization we encountered during our experiments.
We obfuscate the examples due to ethical considerations (e.g., protecting sensitive information and the identity of contributors of vulnerable code).

\vspace*{0.2cm}
\noindent \textbf{Scenario 1: Intellectual property issues.}
Some code models are trained on public repositories on GitHub, without paying much attention to the licensing terms of the code.
Consequently, memorization in code models may cause intellectual property issues, such as violation of open-source licenses.
When the model learns to reproduce specific code snippets from the training data instead of generalizing the underlying concepts,
this leads to the generation of code that closely resembles copyrighted or licensed code, potentially infringing on the intellectual property rights of the original authors.
Listing~\ref{lst:license} shows an output from \texttt{CodeParrot}, which memorizes the original \texttt{read} function verbatim from a public repository that is licensed under Apache License 2.0.
If a developer uses the generated code without proper compliance with the Apache License 2.0 terms, it may violate open-source licenses.

\begin{figure}[h]
    \begin{lstlisting}[caption={An example of code snippets generated by \texttt{CodeParrot}, which is protected by an Apache license.\protect\footnotemark}, captionpos=b, language=Python, label=lst:license, basicstyle=\scriptsize\ttfamily, linewidth=0.45\textwidth]
def read(self, iprot):
  iprot.readStructBegin()
  while True:
    (fname, ftype, fid) = iprot.readFieldBegin()
    if ftype == TType.STOP:
      break
    if fid == 1:
      if ftype == TType.STRING:
        .....
\end{lstlisting}
\end{figure}

\footnotetext{\url{https://github.com/Workiva/frugal}}

\noindent \textbf{Scenario 2: Security vulnerabilities.}
The training data may contain vulnerable and even malicious code.
After being trained on such data, code models might generate malicious code and exploitable bugs, or adopt poor security practices.
If a model memorizes and outputs insecure code to users, it puts users and systems at risk by opening them up to security threats.
Listing~\ref{lst:vulnerable} demonstrates potential security risks related to code model memorization.
The memorized code is susceptible to SQL injection attacks because it directly uses the `\texttt{user\_id}' input in the SQL query without sanitizing or parameterizing it.
An attacker could potentially manipulate the `\texttt{user\_id}' input to execute malicious SQL queries if this generated code is adopted by users.

\begin{figure}[h]
    \begin{lstlisting}[caption={An example of insecure code generated by \texttt{CodeParrot}. The identifiers are substituted and strings are masked due to ethical consideration.}, captionpos=b, language=Python, label=lst:vulnerable, basicstyle=\scriptsize\ttfamily, linewidth=0.45\textwidth]
def fetch_user_data(user_id):
  connection = sqlite3.connect(<masked value>)
  cursor = connection.cursor()
  query = f'SELECT * FROM users WHERE id = {user_id}'
        +  <masked value>
  cursor.execute(query)
\end{lstlisting}
\end{figure}

\noindent \textbf{Scenario 3: Leakage of sensitive information.}
Recent studies show that there exist a vast amount of software secrets in public repositories~\cite{basak2023secretbench, 9794113}, including API keys, passwords, etc.
A code model that tends to memorize data can cause the leakage of sensitive information if it has been exposed to such information during training.
Listing~\ref{lst:sensitive} provides an example that a code model memorizes a hardcoded public IP address, username, and password from its training data.
On the one side, the code model exposes sensitive information, leading to potential security breaches and privacy violations.
On the other side, even the end users do not use the generated information, users may use the program sketch and replace the password with their own, which is still an insecure development practice.

\begin{figure}[h]
    \begin{lstlisting}[caption={An example of source code with sensitive information generated by \texttt{CodeParrot}, including sensitive information like public IP address, username, etc. Identifiers are substituted and strings are masked for ethical consideration.}, captionpos=b, language=Python, label=lst:sensitive, basicstyle=\scriptsize\ttfamily, linewidth=0.45\textwidth]
netowrk_config = {
  'device_type': <masked value>,
  'ip':   <masked value>,
  'username': <masked value>,
  'password': <masked value>
}
network_con = ConnectHandler(**netowrk_config)
print network_con.find_prompt()
\end{lstlisting}
\end{figure}

\subsection{Ethical Consideration}
We emphasize our ethical responsibility by explicitly stating that our goal is to comprehend the memorization phenomenon rather than actively exploiting sensitive information within memorized contents.
Consequently, we avoid utilizing `targeted attacks'~\cite{alkaswan2023targeted} (i.e., aiming to extract specific types of memorization from the training data) to identify or extract sensitive information deliberately.
To further minimize the ethical concerns, we choose two open-source code models that are trained on publicly available datasets rather than models that are commercial.
Throughout the research process, we ensure that we handle data and results in an ethical manner.
For example, in Listing~\ref{lst:vulnerable} and~\ref{lst:sensitive}, we obfuscate identifiers and mask the sensitive information so that the contributors of the vulnerable code cannot be disclosed.

\section{Methodology}
\label{sec:methodology}

This section explains the methodology of our study, including how to sample outputs from code models and how to detect memorization in code models.



\subsection{Generating Outputs from Code Models}
\label{subsec:sampling}


We consider four different strategies to generate outputs from code models: non-prompt generation, temperature-decaying generation, prompt-conditional generation, and two-step generation.

\vspace*{0.2cm}
\noindent
\textbf{Non-Prompt Generation (NPG)}.
We leverage the autoregressive nature of language models to generate outputs without an initial  prompt.
The generation process is described as:

\begin{enumerate}[leftmargin=0.5cm]
	\item Language models usually have a special token to represent the beginning of a sentence (e.g., \texttt{<s>} in GPT-2). We initialize a token sequence with only this start-of-sentence token, denoted by $p = \langle x_1 \rangle$, where $x_1 = \texttt{<s>}$.
	\item For each step $t = 2, 3, \cdots$, we compute the probability distribution of the next token given the current token sequence as input, $P(x_t | p)$ and decide the next token $x_t$. We then append the new token to the token sequence, updating the context for the next iteration: $p \leftarrow \langle p, {x_t} \rangle$.
	\item The above process repeats until a termination criterion is met. In this paper, the termination criterion is specified by the maximal length of output a model generates. Once the termination criterion is met, the generated token sequence is then decoded into human-readable code.
\end{enumerate}

\vspace*{0.2cm} \noindent
\textbf{Temperature-Decaying Generation (TDG)}.
The outputs generated by NPG can often be repetitive and lack of diversity.
To increase the diversity, we can use a larger value for the \textit{temperature} $\tau$.
The temperature value controls the level of diversity in outputs.
Recalling the Equation~\ref{eq:softmax}, the temperature $\tau$ is used to scale the logits before applying the softmax function. Specially, the new distribution is computed as follows:
\begin{equation}
	P(x_{n+1}|p) = \frac{e^{\frac{z_{n+1}}{\tau}}}{\sum_{i=1}^k e^{\frac{z_{i}}{\tau}}}
	\label{eq:temperature}
\end{equation}
When $\tau$ is small, the probability distribution is compressed, which means that the model will be more likely to select the originally most probable tokens.
Otherwise, the model will give less preference to the originally most probable tokens, producing more diverse outputs.
However, maintaining a high temperature increases the probability of generating incoherent code and, as a result, produces less memorization.
Carlini et al.~\cite{carlini21extracting} use the \textit{temperature-decaying} strategy.
Specifically, during the generation process, the temperature gradually decreases from a high value to a low value.
It encourages the model to generate diverse outputs for the first several tokens and then gradually focus on generating coherent outputs.

\vspace*{0.2cm} \noindent
\textbf{Prompt-Conditioned Generation (PCG)}.
This strategy requires designing a prompt for the code model.
Generation from the start token tends to produce many contents that usually appear at the beginning of code files, e.g., license information.
Besides, the users usually send prompts to the code model for completion rather than generating code from the start token.
Therefore, we randomly choose a list of files from unseen testing data and parse the files to extract the function definition statements.
We feed the function definition statements as the prompt to the code model for completion.

\vspace*{0.2cm}
\noindent
\textbf{Two-Step Generation (TSG)}.
In NPG, a code model generates outputs from the start token, which produces a large number of license information and less amount of outputs relevant to the prompt.
Thus, after we identify memorization using NPG, we send the most frequently appearing memorization as the prompt to the code model for completion to explore whether this can drive the model to generate more memorization.

\subsection{Memorization Detection}
\label{subsec:detection}

After generating outputs from the code models, we detect
whether the outputs contain memorization.
Recalling the definition of memorization in Section~\ref{subsec:definition}, a string from model outputs is memorized if this string is also in the training data.
In a previous study analyzing memorization in natural language
models~\cite{carlini21extracting}, the memorization is detected by performing
fuzzy match (e.g., 3-gram fuzzy match) between the generated text and the training data.
However, this strategy is not suitable to code models for the following reasons.
Unlike natural language, source code has a
more structured syntax with specific rules
and conventions, which usually have unique constructs, keywords, and idioms that
often occur together in specific sequences.
A fuzzy match approach might flag these commonly occurring sequences as
memorization despite the fact that they are simply inherent to the language.

We use the concept of code clone to determine whether an output contains memorization.
A {\it Type-1 clone}, also known as an exact clone or an identical clone,
refers to a specific type of code duplication in which two or more code
fragments are exactly the same~\cite{roy2007survey}.
This means that the code
fragments have the same sequence of tokens, including
comments, whitespace, and formatting.
Comparing to other types of clone (e.g., Type-2 clone), Type-1 clone means that a model produces output verbatim from the training data, which is a stronger evidence of memorization.
However, very short Type-1 clones may not be considered memorization and they may not reveal useful information about the model.
For example, the code snippet `\texttt{a = 1}'
may appear in many places in the training data.
As a result, we only consider Type-1 clones that are longer than a threshold number of lines $L$.

Following a study~\cite{code-model-clone-msr22} on analyzing the clones of code generated by deep learning-based code recommenders, we employ \textsc{Simian}~\cite{simian} for identifying Type-1 clones between the generated code and the training data.
Although there are other clone detection tools, some of these tools are not publicly accessible, and others are unable to process incomplete code snippets, which are often produced by language models.
By default, \textsc{Simian} identifies clones that span at least six lines, which is chosen as the threshold $L$ in our study.

\subsection{Memorization Prediction}
\label{subsec:prediction}

While {\it memorization detection} directly compares the output of a code model
with source code in the training data, {\it memorization prediction} infers
whether an output contains memorized content without accessing the training data.
This task assumes that the training data is not accessible;
this may simulate  \textit{adversarial} settings.
For example, a malicious user may generate a large amount of outputs and then
infer which parts are more likely to be memorization.
If such a malicious user can infer memorization correctly, information
regarding the training data (e.g., private code or software secrets) can be
leaked, putting a potential security risk.
According to the study on analyzing memorization in natural language
models~\cite{carlini21extracting}, this paper adopts
the following metrics to infer memorization.

\vspace*{0.2cm} \noindent
\textbf{Perplexity.}
Perplexity~\cite{ppl} measures how well a language model predicts a sample. A lower perplexity value indicates that the model is more confident in its predictions. Intuitively, a model is more confident on the example that it has seen during training. Therefore, a lower perplexity value may indicate that a model retrieves memorization.

\vspace*{0.2cm} \noindent
\textbf{\texttt{PPL}-\texttt{PPL} ratio.}
Language models tend to have low perplexity trivial memorization, e.g., repeated contents such as license information.
From the perspective of risk assessment, exposing code logic leads to higher
risks than exposing trivial memorization.
Imagining that there are two code models:
a small model and a large model, both trained on the same dataset.
As shown in Section~\ref{subsec:rq2},
a large model better memorizes more non-trivial contents than the small model.
It means that on trivial memorization, both two models have small perplexity,
but on non-trivial memorization, the large model has smaller
perplexity than the small model.
Therefore, we use the ratio of perplexity between the large model
and the small model to infer memorization.
We name this metric as \texttt{PPL}-\texttt{PPL} ratio (\texttt{PPL} for perplexity).
The metric is computed as $\frac{log(P_l)}{log(P_s)}$, where $P_s$ is the perplexity computed using the small model and $P_l$ is the perplexity of the large model.
A small ratio indicates that an output is more likely
to contain non-trivial memorization.

\vspace*{0.2cm} \noindent
\textbf{\texttt{PPL}-\texttt{zlib} ratio.}
\texttt{zlib}~\cite{zlibnet} is a data compression library.
The \texttt{zlib} entropy of a text is computed as the number of bits
when the text is compressed with \texttt{zlib} library.
A repetitive input has a small \texttt{zlib} entropy.
We compute the \texttt{PPL}-\texttt{zlib} ratio: $\frac{log(P_v)}{zlib}$ (the ratio of the model perplexity and the \texttt{zlib} entropy).
A small ratio indicates that the output
content is more likely to be memorized but less likely to be repetitive~\cite{carlini21extracting}.

\vspace*{0.2cm} \noindent
\textbf{Average \texttt{PPL}.}
Memorized contents may be surrounded by non-memorized outputs.
Computing the perplexity of an output mixed of memorized
and non-memorized content may not reflect the model's confidence.
As stated in Section~\ref{subsec:detection}, clones spanning over 6 lines are considered as memorization in this paper.
So we apply a sliding window of 6 lines to each output
(moving by one line each time).
We compute the perplexity of each window and then compute
the Average \texttt{PPL} of all windows.
We rank model outputs by the Average \texttt{PPL} in ascending order.


\section{Experiment Setup}
\label{sec:settings}

\subsection{Subject Code Models and Dataset}
\subsubsection{Models for Memorization Analysis}
This paper first analyzes the contents that are exactly memorized from the training data (i.e., Type-1 clones) and the factors that affect such memorization.
To achieve this goal, we investigate two models: \texttt{CodeParrot}
and \texttt{CodeParrot-small}.
The former is based on the \texttt{GPT-2} model with 1.5 billion parameters,
and the latter is a smaller version of \texttt{CodeParrot}, which also uses the
\texttt{GPT-2} architecture but has fewer (110 million) parameters.
Both models are trained on a collection of Python files from scratch.
The two models are available on the HuggingFace model hub.
Table~\ref{table:models} shows the performances of two investigated models and
other code models of similar sizes on
the OpenAI's HumanEval benchmark~\cite{codex}.
Although there are other available code models, this study selects \texttt{CodeParrot} and
\texttt{CodeParrot-small} as the main research subjects for the following reasons.

First, the training data of some models can be either inaccessible or lack
sufficient details, preventing a thorough analysis of
their memorization capabilities.
For example, InCoder~\cite{fried2023incoder} states the source of its training data but does not provide
the exact dataset used for training.
Second, while some models have available data, its extensive dataset, which
includes multiple programming languages, complicates clone
detection-based memorization analysis.
For example, SantaCoder~\cite{santacoder} uses a training set of 3TB and GPT-Neo~\cite{gpt-neo} is trained on a diverse text dataset of 800GB.
Additionally, SantaCoder only comes in one size, making it unsuitable for
analyzing the impact of model size.
Third, commercial models like \textsc{ChatGPT} exhibit superior results but cannot be
analyzed due to the unavailability of their training data.
Attempting to extract training data from these models can be potentially
considered violating their terms of use as well.
Lastly, \texttt{CodeParrot} is a popular model.
Querying ``\texttt{CodeParrot}'' as the keyword on HuggingFace returns 419 models.
In contrast, searching HuggingFace using ``\texttt{codebert}'' and ``\texttt{codet5}'', which are both widely evaluated models in the literature, returns only 126 and 124 models, respectively.\footnote{The result is obtained on 25 July 2023.
}


\begin{table}[!t]
    \centering
    \caption{Model performance on the OpenAI's HumanEval benchmark~\cite{codex}. \textbf{Pass@n} means the chance that a model provides a correct answer within $n$ attempts.}
    \label{table:models}
    \begin{tabular}{@{}lcccc@{}}
    \toprule
    \textbf{Model} & \textbf{Size} & \multicolumn{3}{c}{\textbf{Pass@n}}\\ \cmidrule{3-5}
     & & \textbf{n=1} & \textbf{n=10} & \textbf{n=100} \\ \midrule
     \rowcolor{gray!15} \texttt{CodeParrot} & 1.5B & 3.80\% & 6.57\% & 12.78\% \\
    \texttt{CodeParrot-small} & 110M & 3.58\% & 8.03\% & 14.96\% \\
    \rowcolor{gray!15} \texttt{PolyCoder}  & 160M & 2.13\% & 3.35\% & 4.88\% \\
    \texttt{PolyCoder} & 400M &  2.96\% & 5.29\% & 4.88\% \\
    \rowcolor{gray!15} \texttt{GPT-Neo} & 125M & 0.75\% & 1.88\% & 2.97\% \\
    \texttt{GPT-Neo}  & 1.3B & 4.79\% & 7.47\% & 16.30\% \\
    \bottomrule
    \end{tabular}
\end{table}


The selected models are trained on the \texttt{CodeParrot} dataset,
which was created with the GitHub dataset available via Google's BigQuery.
This dataset contains approximately 22 million Python files and is 180 GB in size.
Some preprocessing steps are conducted to clean the dataset.
After removing whitespace in each file, the developers of \texttt{CodeParrot}
find and remove exact duplicates by computing the hash of each file.
The files whose first 5 lines contain the word `auto-generated'
are removed to remove the potentially generated files.
Then, we follow the preprocessing tasks detailed in the original dataset
repository.
After data cleaning, the processed dataset is split into the training set
and the validation set, which contain approximately 5 million files that are 50GB.

\subsubsection{Analyzing Memorization in Deployed Models}
We also conduct a case study on more code models to analyze whether they can potentially memorize their training data as well.
We select two additional models: \texttt{Incoder}~\cite{fried2023incoder} and \texttt{StarCoder}~\cite{starcoder}.
We consider these models for the following reasons.
(1) The two models are of larger sizes (6B and 15.5B, respectively), popular, and demonstrate strong performance~\cite{fried2023incoder,starcoder}; and
(2) These models have been deployed for practical usage. \texttt{StarCoder} has been deployed as an extension that can be used in IDEs like IntelliJ IDE\footnote{\url{https://plugins.jetbrains.com/plugin/22090-starcoder}}. \texttt{Incoder} is deployed and used within Meta~\cite{murali2023codecompose}. Evaluating these models can help us understand the risks of using code models in practice.


\subsection{Research Questions}

In this study, we answer three research questions to investigate the risk of code models memorizing training data:

\begin{itemize}[leftmargin=0.5cm]
    \item \textbf{RQ1.} \textit{What do code models memorize?}
    \item \textbf{RQ2.} \textit{What factors affect memorization in code models?}
    \item \textbf{RQ3.} \textit{How to infer whether an output contains memorized information?}
  \end{itemize}

The first question aims to understand the extent and nature of information that code models tend to memorize, trying to categorize the memorized contents into different types.
The second question analyze the factors that influence the memorization, which helps us understand the risks associated with code models leaking their training data.
The third question focuses on infering whether an output contains memorized information from its training data.
By addressing these research questions, we seek to get a deeper understanding of the risks associated with code models leaking their training data, along with practical techniques to identify and mitigate such risks. The following paragraphs describe the experiment design for each research question.

\vspace*{0.1cm}
\subsubsection{RQ1. What do code models memorize?}
\label{sec:rq1setup}
~\newline
\noindent
\textbf{Motivation.}
Code models are trained on a large amount of data collected
from various sources,
including both public open-source repositories and private code bases; both of
which are prone to contain sensitive information.
Understanding what code models memorize helps avoid the risks associated with
code models leaking their training data.
For example, if code models can memorize and output information such as
function implementation and even private data,
then the risk of code models leaking their training data is high.

\vspace*{0.2cm}
\noindent \textbf{Experiment Design.}
We make a code model generate a large amount of
source code according to different generation strategies
described in Section~\ref{subsec:sampling}.
For this experiment,
we choose the \texttt{CodeParrot} model as the experiment subject.
For each strategy, we generate 20,000 outputs; each output has a
maximal length of 512 tokens.
Then we analyze the Type-1 clones spanning more than 6 lines that appear in
both the training data and the outputs, i.e., memorized contents.

In order to gain a deeper understanding of the memorization, we design an annotation study to classify the memorization into different categories.
We identify 40,125 unique clones (ranging from 6 to 53 lines) from 20,000 outputs produced by Non-Prompt Generation (NPG).
We use a widely adopted sample size calculator\footnote{\url{https://www.surveysystem.com/sscalc.htm}} with a confidence level of 95\% and a confidence interval of 5 to obtain a statistically representative sample size of 381.
We conduct an open-card sorting study~\cite{10.1145/2568225.2568233}, a well-established technique for generating meaningful groupings of data.
Two authors of this paper discuss with each other to categorize the cards and a senior author resolves the disagreement and merge low-granularity categories into higher-level ones.
We then classify the memorized contents into three main categories: \textit{Documentation}, \textit{Code Logic}, and \textit{Others}, each having 2 to 8 sub-categories.
Each memorized content can contain multiple categories.

\vspace*{0.1cm}
\subsubsection{RQ2. What factors affect memorization in code models?}
\label{subsec:rq2}
\noindent \textbf{Motivation.}
RQ1 shows that code models memorize different types of
contents and that the content of the prompts affects the
model outputs,
motivating us to further investigate the other factors that impact the memorization of code models.
Knowing the relevant factors helps us better understand code models and
highlight potential directions to mitigate memorization.

\vspace*{0.2cm}
\noindent \textbf{Experiment Design.}
We consider the following factors that may affect the memorization of code models.

\begin{itemize}[leftmargin=0.3cm]
  \item \textbf{Model size}: Previous research~\cite{codegen} shows that if two models share the same architecture and the same dataset, the larger model has stronger capacity.
  In this experiment, we compare the memorization power of \texttt{CodeParrot} and its smaller version \texttt{CodeParrot-small}.
  \item \textbf{Top $k$ sampling}: Code models generate the next token by choosing from the top-$k$ most likely tokens. A small (large) $k$ value can generate more fixed (diverse) outputs. We try 4 settings: 5, 10, 20, and 40 to investigate the effect of $k$.
  \item \textbf{Output length}: The length of model outputs (i.e., number of tokens generated) may also impact the memorization of code models. We try 4 settings: 256, 512, 768, and 1024.
  \item \textbf{Number of generated outputs}: If we can query the model for many times and obtain many outputs, the model may expose more memorized information.
  \item \textbf{Occurrences in the training data}: The number of occurrences of a code snippet in the training data can affect the memorization. Intuitively, if the model sees a training example frequently, it may overfit this example and produce the frequently occurring patterns at higher probability.
\end{itemize}

\vspace*{0.1cm}
\subsubsection{RQ3. How to infer whether an output contains memorized information?}
~\newline
\textbf{Motivation}:
For the aforementioned research questions, we assume to have complete access to
the training data in order to analyze memorization.
This assumption is practical when conducting analysis as model developers.
However, in certain adversarial situations, such as when malicious users aim to
extract training data from code models, the attacker typically does not have
knowledge of the training data.
Therefore, we investigate how to infer whether an output from code models
contains memorization without querying the training data.

\noindent \textbf{Experiment Design}:
We use the methods described in Section~\ref{subsec:prediction} to rank the outputs from code models by using four different metrics.
Following the setting in the paper that propose these metrics~\cite{carlini21extracting}, we look into the 100 top-ranking outputs from the ranked lists and compute the ratio of the outputs containing memorized contents.

\subsection{Implementation Details}

We fetch the \texttt{CodeParrot} and \texttt{CodeParrot-small} models
from the HuggingFace model hub and run them an NVIDIA GeForce A5000 GPU with 24 GB of memory.
To implement the temperature-decaying generation,
we use an initial high temperature of 20.0.
Each time the model generates a new token, we decrease the temperature by
1.0 until it reaches 1.0 (i.e., 20 tokens are generated),
after which we keep the temperature at 1.0.
We download the datasets of the two models released
by the authors of \texttt{CodeParrot}.\footnote{\url{https://huggingface.co/datasets/codeparrot/codeparrot-clean}}
As the clone detection tool we use consumes a large amount of time
when analyzing many files, we merge all the files in the training data
and split the merged file into 53 chunks and run clone detection in parallel.
On a machine having AMD EPYC 7643 CPU with 48 cores and 512GB memory,
analyzing memorization for each 20,000 examples in parallel takes one hour.
We encourage researchers with more computational resources to replicate
our study at a larger scale.

\section{Experiment Results}
\label{sec:result}

\subsection{RQ1. What do code models memorize?}

\noindent {\bf Memorization Detection and Categorization.}
Model outputs by different generation strategies include a significant number of
 code fragments memorized from the training data.
Note that, on average, approximately 43\% and 57\% of 20,000 outputs from
\texttt{CodeParrot-small} and \texttt{CodeParrot} contain memorized information, respectively.
Following the procedure described in Section~\ref{sec:rq1setup},
we randomly sample 381 memorized outputs for each generation strategy and categorize them.
Table~\ref{tab:occurrences} shows the total number of occurrences of each category obtained using different generation strategies.
We observe that the \textit{license and copyright}, as well as the \textit{import statements}, are the most frequent categories.
More specifically, the \textit{license and copyright} category appears 216, 150, 288, and 222 times in the outputs generated by NPG, TSG, TDG, and PCG, respectively.
The \textit{import statements} sub-category appears 129, 211, 80, and 112 times, respectively.
The reasons for their higher frequency than other categories are two-fold.
On the one side, such information usually appears at the beginning of a file, which is easy to been seen and memorized.
On the other side, there exist many duplicates of such information (e.g., developers declare the same license in many files).



Each generation strategy shows different distributions of outputs.
For example, we find that TSG tends to generate less license and copyright information (decrease from 216 to 150) and more code logic (increase 239 from 330) than NPG.
The reason is that TSG selects the most frequent memorization found by NPG (which is the license information) as the prompt into code models.
The model will complete the content after the prompt and thus skips the license information to generate more code logic related contents, such as import statements and method definitions.

Temperature has an impact on the diversity of sampled outputs.
For example, TDG drives the models to produce higher portion of license information.
The potential reason is that using a higher temperature diversifies model outputs.
For those contents that are harder to memorize (e.g., code logic), adding more diversity makes the model fail to generate them.
Consequently, most of the outputs generated by TDG are license information
while it generates the least code logic among all generation strategies.
Although the prompts used in PCG already include the license and import
statements, PCG produces a similar number of license and import statements as
NPG. This is because PCG generates code logic to complete the function
first, after which the model generates license and import statements.
This observation contrasts with the conclusion drawn in NLP models that higher temperature values helps the model produce more non-trivial information~\cite{carlini21extracting}, which is the code logic in our case.

\begin{table}[!t]
  \caption{Occurrences of memorization extracted with Non-Prompt Generation (NPG), Temperature Decaying Generation (TDG), Prompt-Condition Generation (PCG), and Two-Step Generation (TSG). The number of annotated outputs for each sampling method is 381. Note that one output may map to multiple categories of memorization.}
  \centering
  \begin{tabular}{lrrrr}
  \toprule
  Category & NPS & TSS & TDS & PCS \\
  \midrule
  \rowcolor{gray!15} Documentation &277 & 247 & 333 & 315 \\
  \hspace{5mm} License and Copyright & 216 & 150 & 288 & 222 \\
  \hspace{5mm} Docstring & 30 & 55 & 10 & 50 \\
  \hspace{5mm} Usage Intruction & 31 & 42 & 35 & 43 \\
  \rowcolor{gray!15} Code Logic & 239 & 330 & 100 & 304 \\
  \hspace{5mm} Import Statements & 129 & 211 & 80 & 112 \\
  \hspace{5mm} Method Definition & 39 & 41 & 2 & 41 \\
  \hspace{5mm} Method Calls & 25 & 30 & 4 & 60 \\
  \hspace{5mm} Class Definition & 17 & 19 & 4 & 18 \\
  \hspace{5mm} Conditional Statements & 14 & 10 & 4 & 33 \\
  \hspace{5mm} Exception Handling & 9 & 11 & 4 & 32 \\
  \hspace{5mm} Testing & 4 & 7 & 2 &  5 \\
  \hspace{5mm} Print Statement and Log & 2 & 1 & 0 & 3 \\
  \rowcolor{gray!15} Others & 60 & 36  & 45 & 57\\
  \hspace{5mm} Configuration & 38 & 23 & 25 & 52 \\
  \hspace{5mm} Unable to Classify & 22 & 13 & 20 &  5 \\
  \bottomrule
  \end{tabular}
  \label{tab:occurrences}
\end{table}

Interestingly, we find that PCG produces more memorization related to code logic than other generation strategies.
For example, the occurrences of \textit{exception handling} increase from 9 in NPG to 32 in PCG, and that of \textit{method calls} increase from 25 to 60.
PCG to some extent simulates the behavior of users when interacting with code models: writing some code like function definition and letting the model complete.
Our results suggest that code models may produce much memorization when interacting with users, posing data leakage risk.

\noindent {\bf Sensitive Information Detection.}
Our experiment also reveals sensitive information memorized
by the code models.
One of the examples is shown in
Listing~\ref{lst:sensitivecase} in which we can identify a private key that
might expose financial accounts.
As discussed in Section~\ref{sec:background}, there might be other types of sensitive information, but we focus on counting IP addresses,
email addresses, and hash keys as they are explicitly identifiable
compared with other types of information such as licensed code or
vulnerabilities.
We use \texttt{detect-secrets},\footnote{\url{https://github.com/Yelp/detect-secrets}} a popular open-source tool to scan three types of sensitive information in the 20,000 outputs generated by NPG.
After removing local IPs and emails containing ``example,'' we find 25 IP addresses, 914 emails, and 25 keys that are exactly same with the information from the training data as well (i.e., they are memorized).\footnote{The detection results of other generation strategies are included in `\texttt{./PPI}' folder in the replication package.}
Our findings warn that such sensitive information requires to be properly handled (e.g., removed) before training code models.


\begin{figure}[!t]
  \begin{lstlisting}[caption={An example of source code with private keys. The identifiers are substituted and strings are masked because of ethical consideration.}, captionpos=b, language=Python, label=lst:sensitivecase, basicstyle=\scriptsize\ttfamily, linewidth=0.45\textwidth]
class SignMessagesTest(BitcoinTestFramework):
  def set_test_params(self):
    self.setup_clean_chain = True
    self.num_nodes = 1
  def run_test(self):
    message = <masked value>
    self.log.info(<masked value>)
    priv_key = <masked value>
\end{lstlisting}
\end{figure}




\begin{tcolorbox}[boxrule=0pt,frame hidden,sharp corners,enhanced,borderline north={1pt}{0pt}{black},borderline south={1pt}{0pt}{black},boxsep=2pt,left=2pt,right=2pt,top=2.5pt,bottom=2pt]
  \textbf{Answers to RQ1}: Code models successfully memorize different types of
content, including both documentation and code logic.
The license and copyright, as well as import statements, are the most frequently appearing memorization. With proper strategies, one can drive the model to produce more
code logic-related memorization.
Code models memorize sensitive information such as private keys and emails.
\end{tcolorbox}

\subsection{RQ2. What factors affect memorization in code models?}
\label{subsec:rq2}

\begin{figure*}[!t]
  \centering
  \begin{subfigure}{0.31\textwidth}
      \centering
      \includegraphics[width=\linewidth]{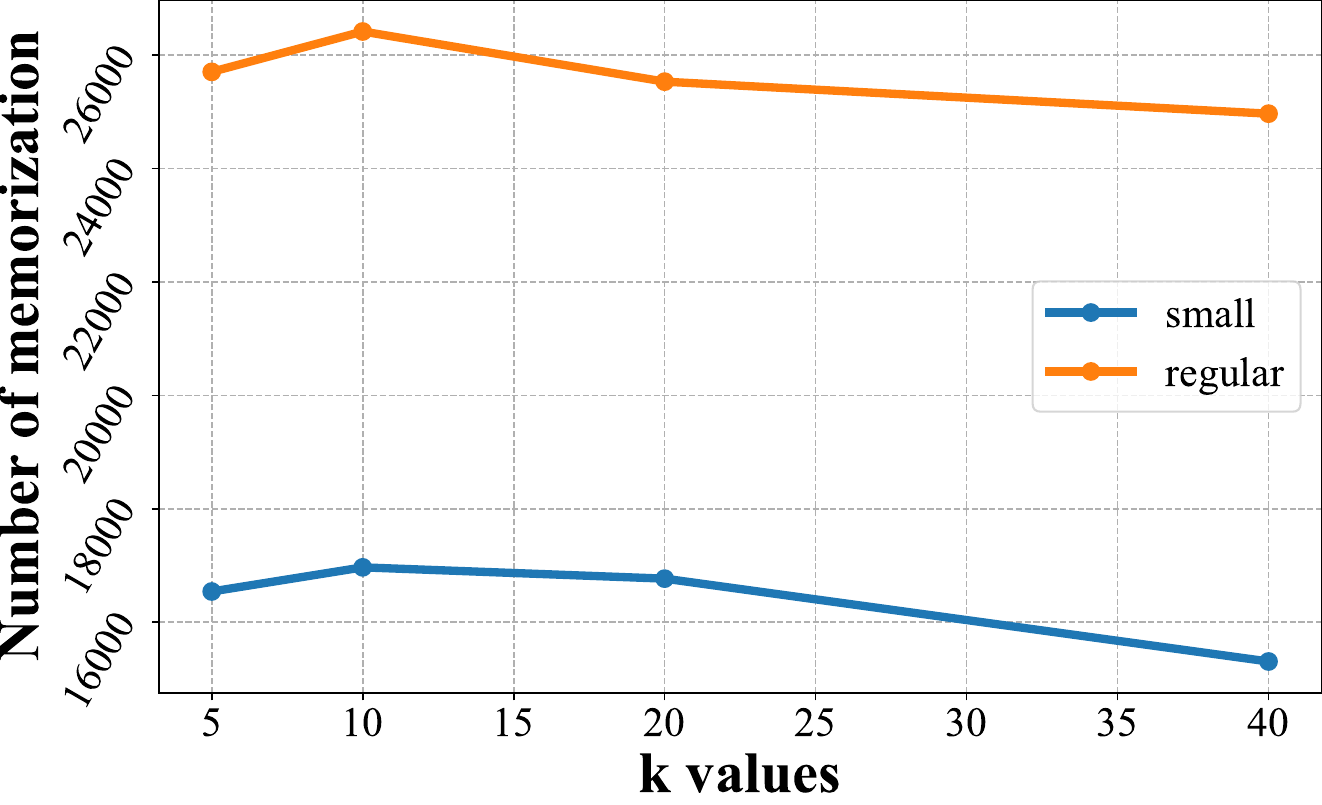}
      \caption{How the $k$ value affects the memorization of code models.}
      \label{fig:values-by-k}
  \end{subfigure}
  \hspace{-0.0cm}
  \begin{subfigure}{0.31\textwidth}
      \centering
      \includegraphics[width=\textwidth]{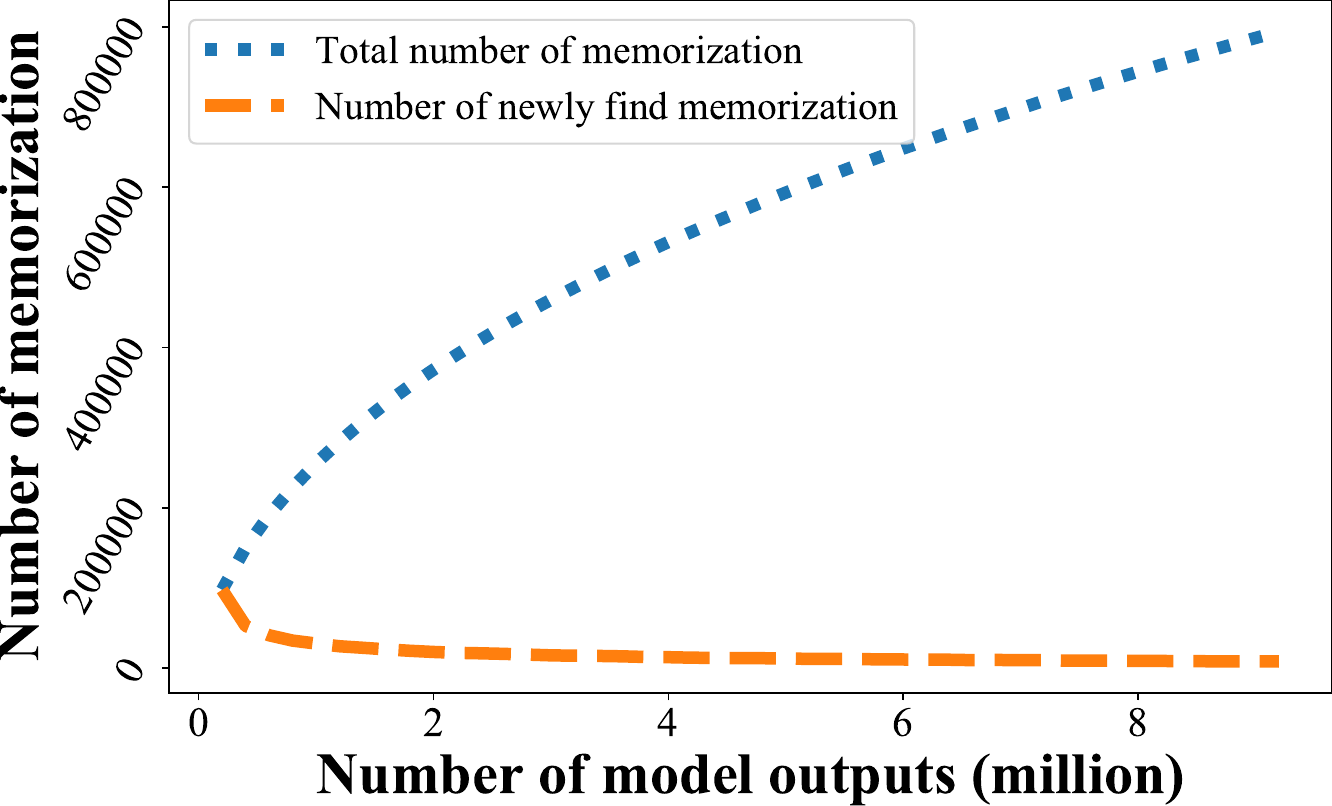}
      \caption{How the total number of outputs affect identified memorization.}
      \label{fig:sample-number}
  \end{subfigure}
  \hspace{-0.0cm}
  \begin{subfigure}{0.31\textwidth}
      \centering
      \includegraphics[width=\textwidth]{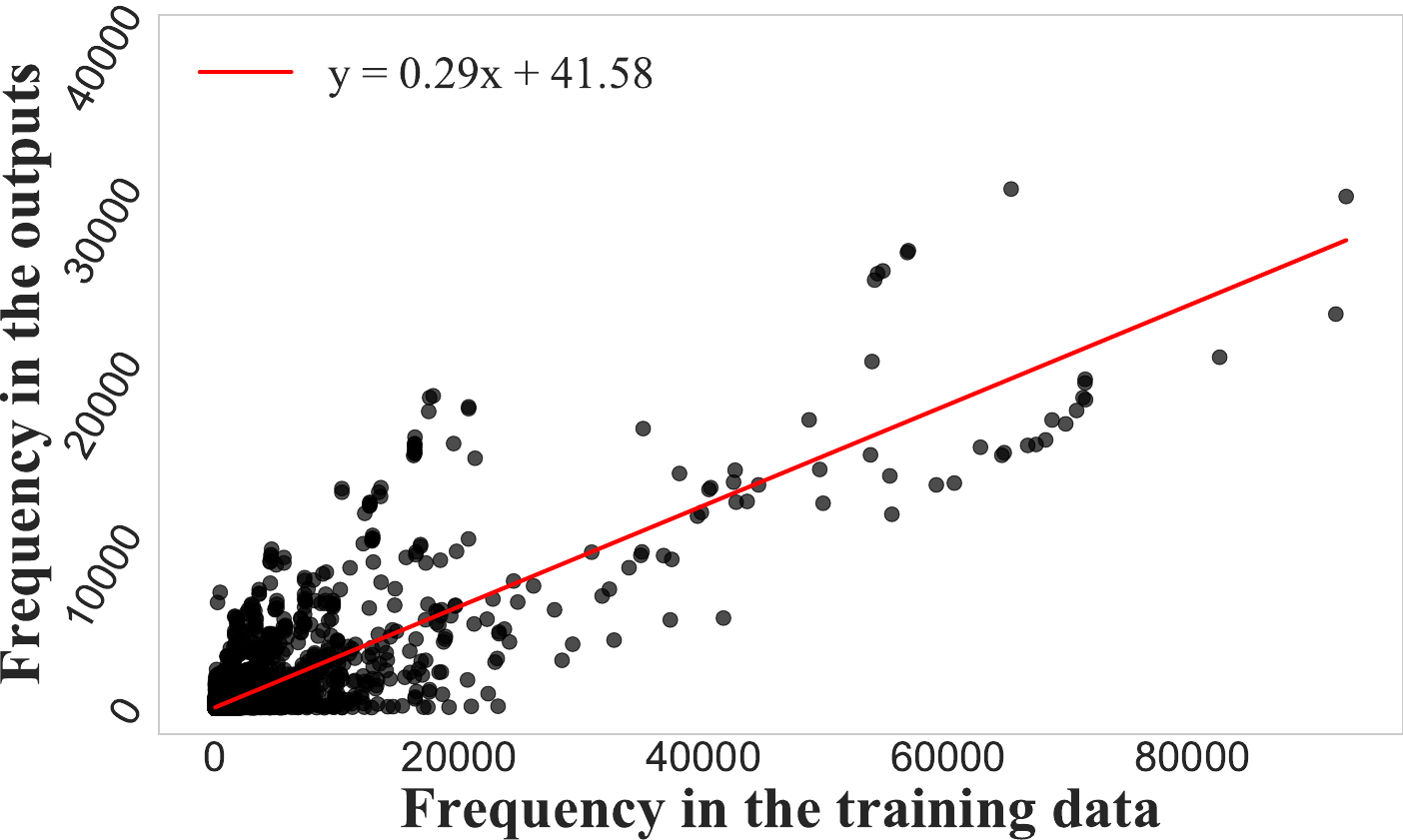}
      \caption{The correlation between the frequency in the training data and outputs.}
      \label{fig:correlation}
  \end{subfigure}
  \caption{Three factors that affect memorization in code models.}
  \label{fig:three_images}
\end{figure*}


\noindent {\bf Model size and top $k$ sampling.}
The model size has a substantial impact on the frequency of memorized outputs
and $k$ parameters affect the results of memorization as well.
Following the procedure described in Section~\ref{subsec:rq2},
we use the Non-Prompt Generation (NPG) to generate 20,000 outputs from \texttt{CodeParrot} and \texttt{CodeParrot-small}, respectively.
Figure~\ref{fig:values-by-k} illustrates the numbers of unique memorization in the outputs of the two models with different $k$ values.
The orange and blue curves represent the results of \texttt{CodeParrot} and \texttt{CodeParrot-small}, respectively.
We observe that curve of the larger model \texttt{CodeParrot} is consistently
above the other curve, suggesting that it memorizes more
contents.
Besides, the larger \texttt{CodeParrot} also memorizes longer contents.
Given the same setting (k=20, output length is 512), \texttt{CodeParrot}
produces memorization with a maximal of 60 lines, while \texttt{CodeParrot-small} only produces memorization with a maximal of 45 lines.

We analyze how the $k$ value affects the memorization of code models.
We try 4 settings of $k$ values: 5, 10, 20, and 40.
The length of each output is set as 256 tokens.
As shown in Figure~\ref{fig:values-by-k}, when $k$ is low (e.g., 5), both models produce less memorization.
As a small $k$ means that a model chooses from a small and relatively fixed set of tokens, which may lead to the generation of memorization with more duplicates (i.e., the number of unique memorization is smaller).
When $k$ increases from 5 to 10, the outputs become more diverse and consequently produce more unique memorized outputs, which can be observed from the figure that the number of unique memorization increases, reaching a peak at $k$=10.
However, when $k$ continues to increase, the outputs become more diverse and differ more from the training data, which leads to a decrease in the number of unique memorization.
As a result, models memorize less alongside $k$ increases from 10 to 40.

\noindent {\bf Output length and the number of outputs.}
Longer outputs can expose more memorized contents according to our experiments.
Given a model, we keep other factors unchanged and vary the maximal number of tokens of the outputs: 256, 512, 768, and 1024.
The results are shown in Table~\ref{tab:tokens}.
When the length of outputs increases, the number of unique memorization also increases.
This trend is consistent for both models on different numbers of outputs.
The impact of output length is larger on the larger model.
Specifically, given the same setting (5,000 examples and increase the output length from 256 to 1024), the number of unique memorization increases by 7,365 (from 6,666 to 14,031 for \texttt{CodeParrot-small}, while increases by 12,785 (from 9,785 to 22,570) for \texttt{CodeParrot}.

Table~\ref{tab:tokens} also shows that when we extract more outputs from the models, the number of unique memorization increases as well.
We let the \texttt{CodeParrot} model to generate 10 million outputs (512 tokens each) and plot the trend of the number of unique memorization with the number of outputs in Figure~\ref{fig:sample-number}.
The blue curve shows the total number of unique memorization, while the orange curve shows the number of newly identified unique memorization by generation additional 200,000 outputs each time.
We see a `diminishing return' pattern: although more memorization can be identified when more outputs are generated, the number of newly identified memorization decreases.
The 10 million outputs contain over 8 million unique memorization that is longer than 6 lines.

\begin{table}[!t]
  \centering
  \caption{Number of unique memorized-code for a given set of parameters: (1) number of tokens and (2) number of outputs.}
  {%
    \begin{tabular}{l| r|cccc}
      \toprule
      Model & \# outputs & 256 & 512 & 768 & 1024 \\
      \midrule
      \multirow{4}{*}{\shortstack{\texttt{CodeParrot-}\\\texttt{small}}}
       & 5,000  & 6,666 & 9,080 & 11,041 & 14,031 \\
       & 10,000 & 10,627 & 14,655 & 17,664 & 22,243 \\
       & 15,000 & 14,015 & 19,444 & 23,863 & 29,133 \\
       & 20,000 & 16,966 & 23,574 & 29,204 & 35,363 \\
      \midrule
      \multirow{4}{*}{\texttt{CodeParrot}}
      & 5,000  & 9,785 & 14,645 & 18,325 & 22,570 \\
      & 10,000 & 16,062 & 24,345 & 32,519 & 37,448 \\
      & 15,000 & 21,560 & 32,666 & 42,853 & 50,127 \\
      & 20,000 & 26,420 & 40,125 & 51,059 & 61,787 \\
      \bottomrule
    \end{tabular}
  }
  \label{tab:tokens}
\end{table}

\noindent {\bf Occurrences in the training data.}
Code fragments more frequently appearing in the training data
are more likely to be generated by the code models.
We count the number of occurrences of each memorized content in the training data and the model outputs and visualize the distribution in Figure~\ref{fig:correlation}.
Spearman's rank correlation test obtains a correlation coefficient of 0.804 ($p < 0.01$), indicating a strong positive correlation between the two variables.
We conduct the Pearson correlation test to measure the linear relationship.
We obtain a correlation coefficient of 0.752 ($p < 0.01$), which indicates a strong positive linear correlation between the two variables.
The finding implies that when the model is exposed to a higher frequency of certain contents in its training data, it tends to memorize it and produces that content more frequently in its outputs, highlighting to importance of conducting data de-duplication before training code models.

\begin{tcolorbox}[boxrule=0pt,frame hidden,sharp corners,enhanced,borderline north={1pt}{0pt}{black},borderline south={1pt}{0pt}{black},boxsep=2pt,left=2pt,right=2pt,top=2.5pt,bottom=2pt]
  \textbf{Answers to RQ2}: We identify five factors affecting memorization: (1) the model size given the same architecture; (2) the hyperparameter $k$ in top-$k$ sampling; (3) the length of generated outputs; (4) the number of generated outputs; (5) the frequency of the code fragment in the training data.
\end{tcolorbox}

\subsection{RQ3. How to infer whether an output contains memorized information?}

\noindent {\bf Prediction performance.}
We use the four metrics described in Section~\ref{subsec:prediction} to rank the outputs generated by the code models and analyze the 100 top-ranking outputs.
Table~\ref{tab:infer} shows the ratio of the top 100 outputs that contain memorization.
We find that generally three out of four metrics can accurately rank the memorized outputs at the top.
Note that, on average, approximately 43\% and 57\% of 20,000 outputs from
\texttt{CodeParrot-small} and \texttt{CodeParrot} contain memorized information.
When the output length is set as 256 tokens, all the top 100 outputs ranked using perplexity, \texttt{PPL}-\texttt{PPL} ratio, and \texttt{PPL}-\texttt{zlib} ratio, contain memorization.
Ranking the outputs using \texttt{PPL}-\texttt{zlib} ratio achieves the best performance, with all the top 100 outputs containing memorization.
The method of \texttt{PPL}-\texttt{PPL} ratio is slightly less effective.
When the output length is 512, its detection accuracy is 79\%.
But in other settings, its accuracy is close to 100\%.
However, the Average \texttt{PPL} metric is less effective in detecting memorization, especially on long outputs.

We further analyze what types of memorization are inferred by these methods.
We conduct another annotation study to count the occurrences of different categories of memorization in the 100 top ranked outputs for \texttt{CodeParrot-small} model (output length is 256).
We find that all the memorized outputs ranked by the perplexity and \texttt{PPL}-\texttt{zlib} Ratio are license information.
However, using \texttt{PPL}-\texttt{PPL} ratio can rank more diverse memorization in the top;  12\% of memorized contents contain code logic.


\begin{tcolorbox}[boxrule=0pt,frame hidden,sharp corners,enhanced,borderline north={1pt}{0pt}{black},borderline south={1pt}{0pt}{black},boxsep=2pt,left=2pt,right=2pt,top=2.5pt,bottom=2pt]
  \textbf{Answers to RQ3}: Three out of four metrics can infer memorization accurately. The \texttt{PPL}-\texttt{PPL} ratio ranks more diverse memorization at the top positions, while memorization highly ranked by the other metrics is mainly license information.
\end{tcolorbox}

\section{Discussion}
\label{sec:discussion}

\subsection{Memorization in Deployed Code Models}
\label{sec:more}

We analyze the memorization of two models that have already been deployed in Meta and IntelliJ IDE: \texttt{Incoder} and \texttt{StarCoder}.
Both \texttt{Incoder} and \texttt{StarCoder} are trained on large-scale datasets (159GB and 3TB) that cover multiple programming languages.
However, the training data of \texttt{Incoder} is not directly available and the magnitude of the training data for \texttt{StarCoder} presents significant challenges when attempting to analyze Type-1 clones.
Note that the authors of the models~\cite{fried2023incoder,starcoder} mention that the training data is curated from GitHub.
Thus, we use the search function provided by GitHub as a proxy and manually confirm the memorization of these models.
If GitHub returns the exact code snippets (spanning over 6 lines), we consider an output contains memorization.

For each model, we extract 20,000 outputs using NPG.
According to the finding from RQ3 that \textit{the \texttt{PPL}-\texttt{PPL} ratio ranks more diverse memorization at the top positions}, we rank the outputs using the \texttt{PPL}-\texttt{PPL} ratio and analyze the top 100 outputs.
Then, we manually search GitHub for each line of code in the top 100 outputs.
We find that 81\% and 75\% of the selected outputs from \texttt{Incoder} and \texttt{StarCoder} are confirmed to be memorized.
Besides, 90.12\% and 65.33\% of memorization from the two models is related to code logic.
The results reveal that there is risk that the two models can memorize and expose their training data.
This fact can be worrying especially when models that trained on private repositories (e.g., internal code in a company) are deployed for public usage.

\begin{figure}[!t]
  \begin{lstlisting}[caption={An example of output from \texttt{StarCoder} that contains a substituted email address: `nnheo@example.com'. The addresses of cryptocurrency are still memorized. We only show the first and last 4 digits to protect privacy.}, captionpos=b, language=Python, label=lst:santacoder, basicstyle=\scriptsize\ttfamily, linewidth=0.45\textwidth]
# Bitcoin Cash (BCH)   qpz3<masked value>5nuk
# Ether (ETH) -        0x84<masked value>c9FB
# Litecoin (LTC) -     Lfk5<masked value>qPvu
# Bitcoin (BTC) -      34L8<masked value>BtTd

# contact :- nnheo@example.com
\end{lstlisting}
\end{figure}

Both \texttt{StarCoder} and \texttt{InCoder} adopt preprocessing methods to remove personal identifiable information and other software secrets in the training data.
They use regular expressions to detect and substitute emails.
The benefit of doing so is reflected in the model outputs.
For example, Listing~\ref{lst:santacoder} shows an output from \texttt{StarCoder}.
We search GitHub for this output; GitHub returns the exact code except a different email address.
We believe this output is memorized from the training data, and it substitutes the email address in the training data.
However, we find that the addresses of cryptocurrency (i.e., the masked values in Listing~\ref{lst:santacoder}) are still memorized.
This highlights the need of appropriately dealing with more types of sensitive information in the training data before training code models and releasing them to the public.


\subsection{Suggestions on Memorization in Code Models}
\label{sec:recommend}

Based on our findings, we provide the following suggestions to deal with memorization in code models.

\vspace*{0.2cm}
\noindent \textbf{1. Data Sources}:
Users from various platforms such as GitHub and Stack Overflow should have the right to explicitly indicate whether their data can be utilized for AI model training.
Otherwise, the code models may memorize the data and expose the data to the other users.
Platforms should support such declarations, allowing users to specify their preferences at different levels of granularity.
For example, a user may allow the main branch of a repository to be used for training, but not the development branch as it may include experimental and unfinished code.
To support the declarations, GitHub can allow users to include a separate file that outlines the rights associated with a repository or provide an option to specify the rights in the account settings.

\vspace*{0.2cm}
\noindent \textbf{2. Data Collection and Processing}:
Data collectors should pay attention to data license information and avoid using data with strict or unclear licenses for training code models.
Appropriate preprocessing of the dataset is necessary, which may include but not limited to detecting and removing personal identifiable information and other software secrets reasonably.
Our study shows that duplicates in the training data are more likely to be memorized.
Allamanis~\cite{duplicate-code} suggests that duplicate code may cause adverse effects on code models.
Removing duplicates in the training data can help reduce memorization.
Additionally, collectors can employ defensive methods to prevent privacy attacks (e.g., data poisoning).

\vspace*{0.2cm}
\noindent \textbf{3. Additional Information for Outputs}:
Code model developers should also offer users sufficient information when they use the model.
For instance, model developers should detect whether an output is likely to be memorized from the training data.
If so, users should be informed of the output's origin and its copyright information.
This not only prevents users from inadvertently violating open-source regulations but also empowers them to make informed decisions about the quality of the output.
For example, users may avoid using an output if it is from a poorly-maintained repository.
Furthermore, our study shows that larger models memorize more contents.
Therefore, the service providers of large code models (e.g., ChatGPT) should take measures to limit the queries to the model to prevent potential privacy breaches.

\begin{table}[!t]
    \centering
    \caption{The performance of different methods in inferring whether an output contains memorized information. The numbers represent the ratio
    of top 100 outputs ranked by each method; i.e., 1.0 indicates that
    the method successfully computes rankings top 100 as memorized source code.}
    \label{tab:infer}
    \begin{tabular}{cccccc}
      \toprule
      & \multicolumn{4}{c}{\textbf{Models}} \\
      \cmidrule(lr){2-5}
      \multirow{-2}{*}{\textbf{Methods}} & \multicolumn{2}{c}{\texttt{CodeParrot}} & \multicolumn{2}{c}{\texttt{CodeParrot-S}} \\
      \cmidrule(lr){2-3} \cmidrule(lr){4-5}
      & t=256 & t=512 & t=256 & t=512 \\
      \midrule
      \rowcolor{gray!15} \textbf{Perplexity} & 1.00 & 0.92 & 1.00 & 1.00 \\
      \textbf{\texttt{PPL}-\texttt{PPL} Ratio} & 1.00 & 0.79 & 0.97 & 0.98 \\
      \rowcolor{gray!15} \textbf{\texttt{PPL}-\texttt{zlib} Ratio} & 1.00 & 1.00 & 1.00 & 1.00 \\
      \textbf{Average \texttt{PPL}} & 0.74 & 0.24 & 0.86 & 0.36 \\
      \bottomrule
    \end{tabular}
  \end{table}

\vspace*{0.2cm}
\noindent \textbf{4. Opt-out Mechanisms}:
Dataset providers should allow users to determine if their data has been included in a dataset.
We also suggest building a tracking system to build a connection between a model and its training data.
Such a tracking system helps identify what models include a specific user's data for training.
As mentioned in Suggestion 1, users should have the right to prevent their code from being used in training and be able to provide their consent.
For data that has already been used, users should have the right to declare or withdraw their consent.
Their data should not only be removed from the dataset but also from the models.
Moreover, it is necessary to develop a certification mechanism to evaluate whether a model has removed the data from its knowledge properly.

\vspace*{0.2cm}
\noindent \textbf{5. Multidisciplinary Collaborations}:
We believe that a multidisciplinary approach is essential to effectively address the outcomes of data memorization in code models.
AI researchers can contribute developing models that minimize memorization while maintaining the model performance. Software engineering researchers can focus on open-source data management and understand user requirements.
Legal professionals provide guidance on copyright and other regulatory requirements to ensure compliance and the protection of users' rights.

\subsection{Threats to Validity}
\label{sec:ttv}
\noindent \textbf{Threats to Internal Validity.}
Internal validity refers to the extent to which a study is free from errors or biases that could invalidate the results.
Our study leverages Type-1 clone detection to identify memorization.
However, the accuracy of clone detection tool may affect our results.
To mitigate this threat, we choose a state-of-the-art clone detection tool, \textsc{Simian}~\cite{simian}, which has been used to analyze code clones between the model output and training data.
Another threat is that using Type-1 clone detection may lead to under estimation of memorization.
For example, a code model may not fully memorize a code snippet and produce a slightly modified version of the code snippet.
In this case, this output is not treated as memorization.

\vspace{0.1cm}
\noindent \textbf{Threats to External Validity.}
External validity refers to the extent to which the results of a study can be generalized to other settings.
This paper evaluates two code models based on the GPT-2 architecture, which is widely used in open-source code models~\cite{fried2023incoder,santacoder}.
However, the results obtained in this paper may not generalize to other models.
To alleviate this threat, we analyze the memorization in two popular code models that are deployed in practice and demonstrate that they also suffer from memorization issues.
We also plan to conduct experiments on more models and datasets in the future.

\section{Related Work}
\label{sec:rel_work}

In this section, we present an overview of studies relevant to this paper, including (1) pre-trained models of code and (2) privacy concerns in AI models.

\subsection{Pre-trained Code Models and Analysis}

Large language models like BERT~\cite{bert,RoBERTa} and GPT~\cite{gpt-2,gpt-3} have excelled in NLP tasks, inspiring pre-trained models for code. 
CodeBERT~\cite{CodeBERT} and a list of similar models (including GraphCodeBERT~\cite{GraphCodeBERT}, CuBERT~\cite{CuBERT}, etc) are developed to produce code embeddings to support downstream tasks like defect prediction.
Many code models leverage the GPT architecture~\cite{gpt-2,gpt-3} to conduct generation tasks.
CodeGPT, which trains the GPT-2 architecture on CodeSearchNet~\cite{husain2019codesearchnet}, is proposed as a baseline model in the CodeXGLUE benchmark~\cite{CodeXGLUE}.
Code models with larger sizes and better performance are also proposed, like InCoder~\cite{fried2023incoder}, CodeGen~\cite{codegen}.
A list of studies~\cite{9609166,10.1145/3533767.3534390,niu2023empirical} have empirically shown the outstanding performance of these models on various tasks.

Researchers have also studied the limitations of code models and threats, highlighting the need to study their trustworthiness~\cite{lo2023trustworthy}.
One of the limits is that a malicious user can generate adversarial examples~\cite{Yefet2020,9825895,alert,Epresentation2021} by adding semantic preserving transformations~\cite{rabin2021generalizability,9678706} to fool code models.
Another threat to the training data is data poisoning.
An attacker can simply inject a small portion of malicious code into the training data to poison the datasets~\cite{codebackdoor,advdoor,you-see,CodePoisoner}; consequently, the obtained code models have backdoors that can be triggered by specific inputs.
Data poison attacks can also harm the performance of code models~\cite{coffee,CoProtector}.

Software engineering researchers have noticed the potential issues brought by memorization in code models. 
Al-Kaswan et al.~\cite{alkaswan2023abuse} discuss the security, privacy, and license implications of memorization in code models.
Our paper conducts the first systematic study of memorization in code models, by categorizing memorized content, analyzing factors that affect memorization, etc.
Rabin et al.~\cite{RABIN2023107066} also evaluate memorization in code models, however, their concept of memorization diverges from ours.
In their paper, a code model learns from noisy dataset by memorizing the noise and fails to generalize to test data, which is more similar to the notion of overfitting.
Our paper investigates to what extent code models memorize and output the training data verbatim.

\subsection{Privacy Concerns in AI Models}
Henderson et al.~\cite{henderson2018ethical} discuss some ethical challenges in data-driven dialogue systems, one of which is the potential privacy violation.
Carlini et al.~\cite{carlini2019secretsharer} evaluate unintended memorization in Google's Smart Compose that can complete emails. 
Feldman~\cite{feldman2020learning} uses the long-tail theory to explain the memorization behavior of deep learning models.
Zhu et al.~\cite{zhu2021deepmemory} analyzes memorization behavior for a LSTM-based neural language model.
One important privacy concern related to memorization is the feasibility of data extraction attack, which aims to extract the training data from the model.
An important privacy threat related to memorization is the data extraction attack, which aims to extract the training data from the model.
Carlini et al.~\cite{carlini21extracting} extract around 600 examples of memorization from the GPT-2 model like URLs, phone numbers, etc.
Al-Kaswan et al.~\cite{alkaswan2023targeted} propose a targeted attack on extracting data from the GPT-Neo model.
To the best of our knowledge, our paper presents the first empirical study of memorization in large pre-trained code models.
Another important privacy concern is the membership inference attack (MIA), which aims to infer whether a specific data sample is used to train a model.
Shokri et al.~\cite{7958568} propose a black-box MIA method on machine learning-based classification models.
Hisamoto et al~\cite{hisamoto-etal-2020-membership} operate MIA on machine translation systems. 
Chen et al.~\cite{gan-leak} produces a taxonomy of membership inference attacks against various generative models. 
Mireshghallah et al.~\cite{quantifying} use MIA to quantify the privacy risk of masked language models like BERT. 
Researchers also propose defensive method to mitigate the risks by MIA.
For example, Tang et al.~\cite{tang2022mitigating} propose to use model ensemble to mitigate the MIA risk.
Nasr et al.~\cite{Nasr-defense} leverage adversarial regularization to enhance membership privacy.

\section{Conclusion and Future Work}
\label{sec:conclusion}

This paper conducts a comprehensive study to examine the memorization in large pre-trained code models.
We develop a taxonomy for memorized contents, consisting of 3 primary categories and 14 subcategories.
The study uncovers several key factors affecting memorization: the model size, the length of outputs, etc.
We also discover a strong positive correlation between the frequency of an output's appearance in the training data and that in model outputs.
This suggests that eliminating duplicates in the training data could potentially reduce memorization.
Furthermore, we identify effective metrics that accurately determine whether an output contains memorized contents and offer recommendations on how to address the issue of memorization in code models.
Additionally, we evaluate memorization in two popular models that are already deployed.

In future work, we plan to study larger code models and more programming languages.
We also plan to explore different strategies to mitigate memorization in code models.

\begin{tcolorbox}[colback=white, colframe=black, boxrule=0.4pt]
    The replication package is provided for replication at \textbf{\url{https://github.com/yangzhou6666/Privacy-in-Code-Models}}, which should not be used for malicious purposes like conducting data extraction attacks.
\end{tcolorbox}

\section*{Acknowledgment}
This research / project is supported by the National Research Foundation, under its Investigatorship Grant (NRF-NRFI08-2022-0002). Any opinions, findings and conclusions or recommendations expressed in this material are those of the author(s) and do not reflect the views of National Research Foundation, Singapore. This work was supported by the National Research Foundation of Korea (NRF) grant funded by the Korea government (MSIT) (No. 2021R1A5A1021944 and 2021R1I1A3048013). Additionally, the research was supported by Kyungpook National University Research Fund, 2020.

\balance
\bibliographystyle{ACM-Reference-Format}
\bibliography{../reference}

\vspace{12pt}

\end{document}